\documentclass[prl,twocolumn,amsmath,superscriptaddress,amssymb]{revtex4}

\usepackage{graphicx}
\usepackage{dcolumn}
\usepackage{bm}
\usepackage{amssymb}
\usepackage{amsmath}
\usepackage{wasysym}
\usepackage{color}
\usepackage{times}

\usepackage{placeins}
\usepackage{epstopdf}
\usepackage{upgreek}
\usepackage{hyperref}

\usepackage[normalem]{ulem}

\makeatletter
\renewcommand*\env@matrix[1][c]{\hskip -\arraycolsep
  \let\@ifnextchar\new@ifnextchar
  \array{*\c@MaxMatrixCols #1}}
\makeatother
\begin{document}

\title{Surface States and Rashba-type Spin Polarization in Antiferromagnetic MnBi$_2$Te$_4$(0001)}

\author{R. C. Vidal}\affiliation{Experimentelle Physik VII and W\"urzburg-Dresden Cluster of Excellence ct.qmat, Universit\"at W\"urzburg, Am Hubland, D-97074 W\"urzburg, Germany}
\author{H. Bentmann}\email{Hendrik.Bentmann@physik.uni-wuerzburg.de}\affiliation{Experimentelle Physik VII and W\"urzburg-Dresden Cluster of Excellence ct.qmat, Universit\"at W\"urzburg, Am Hubland, D-97074 W\"urzburg, Germany}
\author{T. R. F. Peixoto}\affiliation{Experimentelle Physik VII and W\"urzburg-Dresden Cluster of Excellence ct.qmat, Universit\"at W\"urzburg, Am Hubland, D-97074 W\"urzburg, Germany}
\author{A. Zeugner}\affiliation{Technische Universit\"at Dresden, Faculty of Chemistry and Food Chemistry, Helmholtzstra{\ss}e 10, D-01069 Dresden, Germany}
\author{S. Moser}\affiliation{Advanced Light Source, Lawrence Berkeley National Laboratory, Berkeley, CA 94720, USA}\affiliation{Experimentelle Physik IV and W\"urzburg-Dresden Cluster of Excellence ct.qmat, Universit\"at W\"urzburg, Am Hubland, D-97074 W\"urzburg, Germany}
\author{C. H. Min}\affiliation{Experimentelle Physik VII and W\"urzburg-Dresden Cluster of Excellence ct.qmat, Universit\"at W\"urzburg, Am Hubland, D-97074 W\"urzburg, Germany}
\author{S. Schatz}\affiliation{Experimentelle Physik VII and W\"urzburg-Dresden Cluster of Excellence ct.qmat, Universit\"at W\"urzburg, Am Hubland, D-97074 W\"urzburg, Germany}
\author{K. Ki{\ss}ner}\affiliation{Experimentelle Physik VII and W\"urzburg-Dresden Cluster of Excellence ct.qmat, Universit\"at W\"urzburg, Am Hubland, D-97074 W\"urzburg, Germany}
\author{M. \"Unzelmann}\affiliation{Experimentelle Physik VII and W\"urzburg-Dresden Cluster of Excellence ct.qmat, Universit\"at W\"urzburg, Am Hubland, D-97074 W\"urzburg, Germany}
\author{C. I. Fornari}\affiliation{Experimentelle Physik VII and W\"urzburg-Dresden Cluster of Excellence ct.qmat, Universit\"at W\"urzburg, Am Hubland, D-97074 W\"urzburg, Germany}
\author{H. B. Vasili}\affiliation{ALBA Synchrotron Light Source, E-08290 Cerdanyola del Valles, Spain}
\author{M. Valvidares}\affiliation{ALBA Synchrotron Light Source, E-08290 Cerdanyola del Valles, Spain}
\author{K. Sakamoto}\affiliation{Department of Material and Life Science, Osaka University, Osaka 565-0871, Japan}
\author{D. Mondal}\affiliation{Istituto Officina dei Materiali (IOM)-CNR, Laboratorio TASC, Trieste 34149, Italy}
 \affiliation{ICTP (International Centre for Theoretical Physics (ICTP), Strada Costiera 11, I-34100 Trieste, Italy}
\author{J. Fujii}\affiliation{Istituto Officina dei Materiali (IOM)-CNR, Laboratorio TASC, Trieste 34149, Italy}
\author{I. Vobornik}\affiliation{Istituto Officina dei Materiali (IOM)-CNR, Laboratorio TASC, Trieste 34149, Italy}
\author{S. Jung}\affiliation{Diamond Light Source, Harwell Campus, Didcot OX11 0DE, United Kingdom}
\author{C. Cacho}\affiliation{Diamond Light Source, Harwell Campus, Didcot OX11 0DE, United Kingdom}
\author{T. K. Kim}\affiliation{Diamond Light Source, Harwell Campus, Didcot OX11 0DE, United Kingdom}
\author{R. J. Koch}\affiliation{Advanced Light Source, Lawrence Berkeley National Laboratory, Berkeley, CA 94720, USA}
\author{C. Jozwiak}\affiliation{Advanced Light Source, Lawrence Berkeley National Laboratory, Berkeley, CA 94720, USA}
\author{A. Bostwick}\affiliation{Advanced Light Source, Lawrence Berkeley National Laboratory, Berkeley, CA 94720, USA}
\author{J. D. Denlinger}\affiliation{Advanced Light Source, Lawrence Berkeley National Laboratory, Berkeley, CA 94720, USA}
\author{E. Rotenberg}\affiliation{Advanced Light Source, Lawrence Berkeley National Laboratory, Berkeley, CA 94720, USA}
\author{J. Buck}\affiliation{
DESY Photon Science, Deutsches Elektronen-Synchrotron, Notkestrasse 85, 22607 Hamburg, Germany}
\author{M. Hoesch}\affiliation{
DESY Photon Science, Deutsches Elektronen-Synchrotron, Notkestrasse 85, 22607 Hamburg, Germany}
\author{F. Diekmann}\affiliation{
Institut f\"ur Experimentelle und Angewandte Physik, Christian-Albrechts-Universit\"at zu Kiel, 24098 Kiel, Germany}
\author{S. Rohlf}\affiliation{
Institut f\"ur Experimentelle und Angewandte Physik, Christian-Albrechts-Universit\"at zu Kiel, 24098 Kiel, Germany}
\author{M. Kall\"ane}\affiliation{
Institut f\"ur Experimentelle und Angewandte Physik, Christian-Albrechts-Universit\"at zu Kiel, 24098 Kiel, Germany}
\affiliation{Ruprecht Haensel Laboratory, Kiel University and DESY, Germany}
\author{K. Rossnagel}\affiliation{
DESY Photon Science, Deutsches Elektronen-Synchrotron, Notkestrasse 85, 22607 Hamburg, Germany}\affiliation{
Institut f\"ur Experimentelle und Angewandte Physik, Christian-Albrechts-Universit\"at zu Kiel, 24098 Kiel, Germany}
\affiliation{Ruprecht Haensel Laboratory, Kiel University and DESY, Germany}
\author{M.\,M. Otrokov}
\affiliation{Centro de F\'{i}sica de Materiales (CFM-MPC), Centro Mixto CSIC-UPV/EHU,  20018 Donostia-San Sebasti\'{a}n, Basque Country, Spain}
\affiliation{IKERBASQUE, Basque Foundation for Science, 48011 Bilbao, Basque Country, Spain}
\affiliation{Donostia International Physics Center (DIPC), 20018 Donostia-San Sebasti\'{a}n, Basque Country, Spain}
\author{E.\,V.~Chulkov}
\affiliation{Donostia International Physics Center (DIPC), 20018 Donostia-San Sebasti\'{a}n, Basque Country, Spain}
\affiliation{Departamento de F\'{\i}sica de Materiales UPV/EHU, 20080 Donostia-San Sebasti\'{a}n, Basque Country, Spain}
\affiliation{Tomsk State University, 634050 Tomsk, Russia}
\affiliation{Saint Petersburg State University, 198504 Saint Petersburg, Russia}
\author{M. Ruck}\affiliation{Technische Universit\"at Dresden, Faculty of Chemistry and Food Chemistry, Helmholtzstra{\ss}e 10, D-01069 Dresden, Germany}\affiliation{W\"urzburg-Dresden Cluster of Excellence ct.qmat, Technische Universit\"at Dresden, Helmholtzstra{\ss}e 10, D-01069 Dresden, Germany}
\author{A. Isaeva}
\affiliation{W\"urzburg-Dresden Cluster of Excellence ct.qmat, Technische Universit\"at Dresden, Helmholtzstra{\ss}e 10, D-01069 Dresden, Germany}\affiliation{Leibniz IFW Dresden, Institute for Solid State Research, Helmholtzstraße 20, D-01069 Dresden, Germany}\affiliation{Institut f\"ur Festk\"orper- und Materialphysik, Technische Universit\"at Dresden, D-01062 Dresden, Germany}
\author{F. Reinert}\affiliation{Experimentelle Physik VII and W\"urzburg-Dresden Cluster of Excellence ct.qmat, Universit\"at W\"urzburg, Am Hubland, D-97074 W\"urzburg, Germany}
\date{\today}

\begin{abstract}
The layered van der Waals antiferromagnet MnBi$_2$Te$_4$ has been predicted to combine the band ordering of archetypical topological insulators like Bi$_2$Te$_3$ with the magnetism of Mn, making this material a viable candidate for the realization of various magnetic topological states. We have systematically investigated the surface electronic structure of MnBi$_2$Te$_4$(0001) single crystals by use of spin- and angle-resolved photoelectron spectroscopy (ARPES) experiments. In line with theoretical predictions, the results reveal a surface state in the bulk band gap and they provide evidence for the influence of exchange interaction and spin-orbit coupling on the surface electronic structure. 
\end{abstract}
\maketitle

The hallmark of a topological insulator is a single spin-polarized Dirac cone at the surface which is protected by time reversal-symmetry and originates from a band inversion in the bulk \cite{konig:07.2,Hasan:10.11}. Notably, breaking time-reversal symmetry by magnetic order does not necessarily destroy the non-trivial topology but instead may drive the system into another topological phase. One example is the quantum anomalous Hall (QAH) state that has been observed in magnetically doped topological insulators \cite{chang:13}. The QAH state, in turn, may form the basis for yet more exotic electronic states, such as axion insulators \cite{mogi:17,Xiao:18} and chiral Majorana fermions \cite{He:17}. Another example is the antiferromagnetic topological insulator state which is protected by a combination of time-reversal and lattice translational symmetries \cite{Moore:10}.

\begin{figure*}
\includegraphics[width=0.95\linewidth]{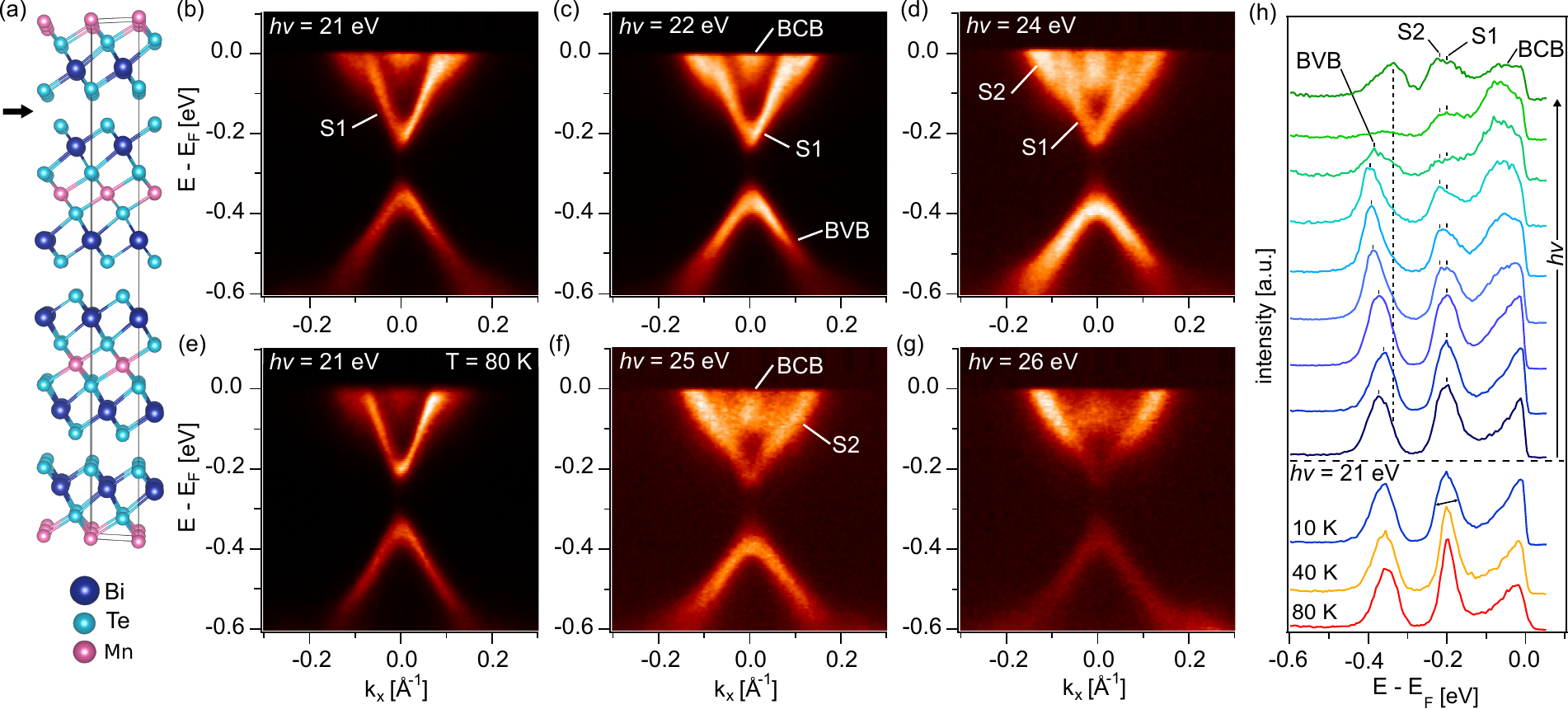}
\caption{(color online) (a) Crystal structure of MnBi$_2$Te$_4$ along one unit cell (black lines). The arrow indicates a van der Waals-gap between two septuple layers and thus a natural cleavage plane. (b-d),(f-g) ARPES data along the $\bar{\Gamma}\bar{\mathrm{M}}$-direction at photon energies from 21$\,$eV to $26\,$eV. The excitation energy dependent evolution of bulk conduction (BCB) and valence band (BVB) states can be seen as well as the states S1 and S2 with two-dimensional character. All data sets were taken at $\mathrm{T} = 10\,$K. (e) ARPES data set taken at $\mathrm{T} = 80\,\mathrm{K}$ and $hv = 21\,$eV.  (h) Energy distribution curves (EDC) at the $\bar{\Gamma}-$point and $hv = 20-28\,\mathrm{eV}$. Markers trace the energy positions of the features BVB, S1 and S2. The dashed line indicates a weak shoulder at the low-binding-energy-side of the BVB peak.}
\label{fig1}
\end{figure*}

Magnetic order in a topological insulator has mainly been achieved by doping with 3$d$ impurities \cite{chen:10,chang:13}, which however inevitably gives rise to increased disorder. By contrast, the layered van der Waals material MnBi$_2$Te$_4$ \cite{Lee:13,eremeev:17} has recently been proposed to realize an intrinsic magnetic topological insulator \cite{otrokov:18,zhang:18,li:18,gong:18}, i.e. a compound that features magnetic order and a topologically non-trivial bulk band structure at the same time. MnBi$_2$Te$_4$ is isostructural to the known topological insulators GeBi$_2$Te$_4$ \cite{Neupane:12,Okamoto:12,arita:14} and PbBi$_2$Te$_4$ \cite{Souma:12}, but the substitution of Ge or Pb by Mn introduces local magnetic moments which order antiferromagnetically below $\sim$24~K \cite{Isaeva:18,otrokov:18,lee:18,Yan:19}. The resulting interplay of magnetic order and topology in a single compound has been proposed as a promising platform for the realization of a variety of magnetic topological states \cite{otrokov:18,zhang:18,li:18,gong:18,rienks:18,hirahara:17,otrokov:17,otrokov:17_2,Otrokov:19}. First-principles calculations predict the presence of a single massive Dirac fermion on the MnBi$_2$Te$_4$(0001) surface in the presence of antiferromagnetic order \cite{otrokov:18,zhang:18,li:18}. Although, angle-resolved photoemission experiments for this surface have been reported recently \cite{otrokov:18,gong:18,lee:18,Wang:19}, a comprehensive understanding of its electronic structure is still lacking.     

In this work we present a systematic investigation of the MnBi$_2$Te$_4$(0001) surface. Employing angle-resolved photoelectron spectroscopy (ARPES) we observe a surface state in the bulk band gap in agreement with theoretical predictions \cite{otrokov:18,zhang:18,li:18}. Temperature-dependent ARPES measurements and resonant photoemission at the Mn $L$-edge are used to address the influence of the magnetic-exchange-split Mn 3$d$ on the electronic structure. By use of spin-resolved ARPES we observe a Rashba-type spin texture in the surface electronic structure, evidencing a strong impact of spin-orbit interaction on the electronic states at the surface.

\begin{figure}
\includegraphics[width=0.9\linewidth]{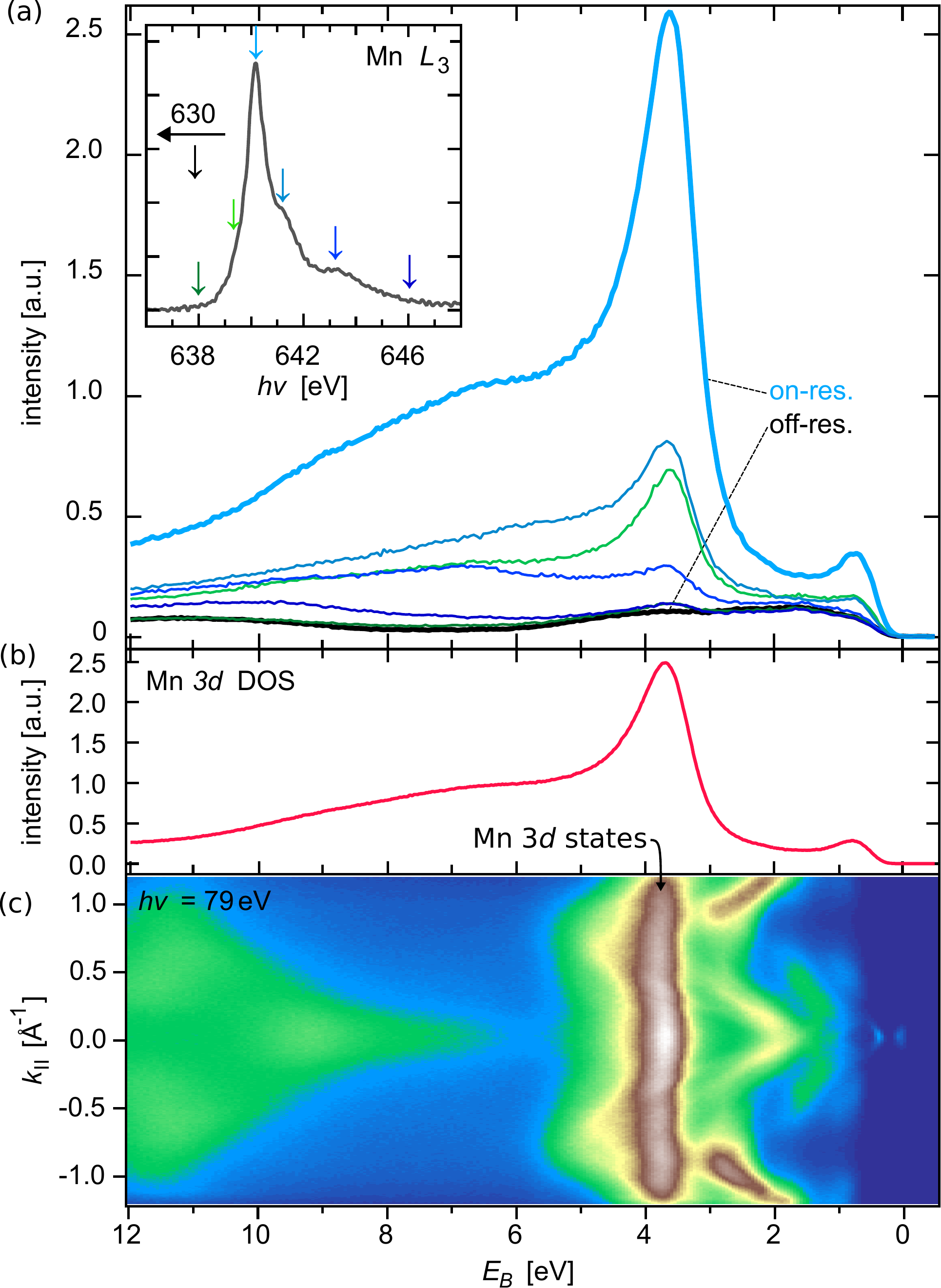}
\caption{(color online) (a) Valence band spectra for MnBi$_2$Te$_4$(0001) obtained by resonant excitation at photon energies $h\nu$ near the Mn $2p \rightarrow 3d$ absorption edge. The inset shows the Mn $L_3$ absorption edge with colored arrows indicating the corresponding excitation energies. (b) Difference of the spectra measured under on- (light blue) and off-resonant (black) conditions, showing the contribution of the Mn $3d$ states to the valence band. (c) Angle-resolved valence band spectrum measured with $h\nu = 79\,$eV.}
\label{fig2}
\end{figure}

ARPES measurements were performed at beamline I05 of the Diamond Light Source (UK) [Fig. 1] with an energy resolution $<10\,$meV as well as at the MAESTRO [Fig.~S2 of the supplemental material \cite{Supp}, which also includes references to \cite{Kuiper:93,Alders:98,abbate:92}] and Merlin [Fig. \ref{fig2}(c)] endstations of the Advanced Light Source (USA) with energy resolutions of $<10\,$meV and $<20\,$meV, respectively. Spin-resolved ARPES measurements were performed at the APE beamline of the Elettra synchrotron by use of a Scienta DA30 hemispherical analyser combined with a spin polarimeter based on very low-energy electron diffraction (Sherman function $S =$~0.3). Resonant soft X-ray photoemission data and Mn 2$p$ core-level spectra were collected at the ASPHERE III endstation at beamline P04 of PETRA III (Germany) with an energy resolution of ca.~25~meV. Platelet-like MnBi$_2$Te$_4$ single crystals were obtained via an optimized crystal-growth procedure and characterized by X-ray diffraction and energy-dispersive X-ray spectroscopy, as described in Ref.~\cite{Isaeva:18}.

MnBi$_2$Te$_4$ crystallizes in a trigonal lattice (the ordered GeAs$_2$Te$_4$ structure type) with septuple [Te-Bi-Te-Mn-Te-Bi-Te] layers stacked in the {\it ABC} fashion \cite{Lee:13}. The septuple layers are separated by a van der Waals gap, as shown in Fig.~\ref{fig1}(a). Our X-ray single-crystal study confirms the structure motif plus a certain degree of Mn/Bi antisite intermixing in both cation positions \cite{Isaeva:18}. Mn atoms are ordered in definite crystallographic sites in a periodic crystal lattice in contrast to doped topological insulators. Cleavage of single cyrstals exposes well-ordered (0001) surfaces suitable for surface sensitive experiments. From an inspection of the Mn 2$p$ core-level line shapes in X-ray photoemission and absorption (XAS and XPS) in Fig.~S1 and S2 we infer a 3$d^{5}$ configuration of the Mn ions (Mn$^{2+}$) \cite{Kurata:93,qiao:13}. Magnetic susceptibility measurements of bulk MnBi$_2$Te$_4$ imply an antiferromagnetic ground state below $T_N =$~24~K \cite{Isaeva:18,otrokov:18,lee:18}. Furthermore, they indicate an out-of-plane orientation of the magnetic moments with ferromagnetic intralayer coupling and antiferromagnetic interlayer coupling, as confirmed by neutron powder diffraction experiments \cite{Yan:19}. This is in agreement with first-principles calculations \cite{li:18,otrokov:18} and in line with our previous XMCD and XMLD measurements \cite{Isaeva:18,otrokov:18} (the data are reproduced in Fig.~S1 of the supplemental material \cite{Supp}).  

Figure~1 presents our ARPES data for MnBi$_2$Te$_4$(0001) near the Fermi level $E_F$. Qualitatively, the electronic structure resembles the ones of the non-magnetic parent compound Bi$_2$Te$_3$(0001) \cite{chen:09} and of isostructural GeBi$_2$Te$_4$(0001) \cite{Neupane:12,Okamoto:12,arita:14} , which are established topological insulators. However, for all used photon energies the spectra for MnBi$_2$Te$_4$(0001) show a gap at the $\bar{\Gamma}$-point. Similar gaps were observed for the topological insulator Bi$_2$Se$_3$(0001) after dilute doping with magnetic Fe \cite{chen:10} and Mn \cite{sanchez:16} impurities. Previous first-principles calculations for for antiferromagnetic MnBi$_2$Te$_4$(0001) predict a gap in the surface electronic structure of ca. 50-100 meV \cite{otrokov:18,zhang:18,li:18}.  

We now analyze the ARPES data in Fig.~\ref{fig1} in more detail. A weak $k_z$ dispersion is found for the main valence-band feature BVB indicating its bulk origin. This is also seen in the energy distribution curves (EDC) at $\bar{\Gamma}$, shown in Fig.~\ref{fig1}(h). A closer inspection of the EDC suggests the presence of another valence-band-related feature, visible as a shoulder at the low binding energy side of the main BVB peak, which could be related to an additional state of surface origin. This interpretation is supported by the supplementary ARPES data set in Fig. S3, where a linearly dispersive feature can be discerned at the top of the valence band. First-principles calculations indeed predict a superposition of bulk and surface states at the valence band maximum \cite{otrokov:18}. Close to the Fermi level we find another state with an apparent $k_z$ dependence, BCB, which we attribute to bulk conduction-band states. Its rather blurred spectral appearance has been observed similarly in n-doped topological insulators like Bi$_2$Se$_3$ \cite{xia:09} and GeBi$_2$Te$_4$ \cite{arita:14}.

Furthermore, we observe the two states S1 and S2 whose spectral weight changes strongly with photon energy $h\nu$ and, particularly for S2, also with binding energy. Within experimental uncertainty we find no significant variations in the dispersion of these states with $h\nu$ in the studied energy range, in clear contrast to the states BVB and BCB. We therefore assign a 2D-like character to S1 and S2. The state S1 has a band minimum at -0.2~eV and the state S2 ca.~20 meV lower, as seen in the EDC in Fig.~\ref{fig1}(i). The dispersion of S2, furthermore, shows a characteristic change in slope at approximately -0.15~eV, where also its intensity in Fig.~\ref{fig1}(f)-(g) quickly varies with binding energy.  

Overall, our measurements are in reasonable agreement with first-principles calculation of the (0001) surface of MnBi$_2$Te$_4$ assuming ferromagnetic intralayer and antiferromagnetic interlayer coupling \cite{otrokov:18,zhang:18,li:18}, i.e. in accordance with the experimentally determined magnetic state \cite{Isaeva:18,otrokov:18,lee:18,Yan:19}. In particular, surface-projected calculations for a semi-infinite half space reveal the presence of two 2D states with electron-like dispersion with band minima at approximately -0.02 eV and 0.08 eV \cite{zhang:18}. Taking into account the n-doping of our samples, these states appear to qualitatively match with the states S1 and S2 observed in Fig.~\ref{fig1}. By direct comparison to the calculations in Ref.~\cite{zhang:18}, we can thus relate the state S2 to the predicted gapped topological surface state (TSS). This TSS arises from the non-trivial $Z_2$ topology of MnBi$_2$Te$_4$ in the antiferromagnetic state and the gap has been attributed to ferromagnetic intralayer ordering \cite{otrokov:18,zhang:18,li:18}. The state S1 may be viewed as a more bulk-like state with surface-resonance character. Starting approximately 0.15 eV above the band minimum of this S1 state, the calculation shows a continuum of bulk states in good agreement with the blurred feature BCB we observe experimentally \cite{zhang:18}. Furthermore, ca. 0.17 eV below this state the calculation predicts the topmost valence band states which is again in reasonable agreement with our ARPES data in Fig.~\ref{fig1}.        

The strong cross-section variations of S1 and S2 are also evidenced by very recently reported ARPES experiments performed at low excitation energies of ca. 6-7 eV \cite{Wang:19,Hao:19,Chen:19,Swatek:19}. Here the photoemission intensity behavior near the $\bar{\Gamma}$ point is considerably different from our data in Fig.~\ref{fig1}, and the spectral gap has even been interpreted as fully closed \cite{Hao:19}. The latter has been attributed to a possible absence of magnetic order near the surface even significantly below $T_N$ \cite{Hao:19}. We note that for the samples studied in the present work \cite{Isaeva:18} and also for the closely related compound MnBi$_4$Te$_7$ \cite{Vidal:19} our surface-sensitive XMLD and XMCD measurements provide no indication for a strong change of the magnetic state near the surface [see Fig.~S1].  

To study the influence of the antiferromagnetic order on the surface electronic structure we next consider temperature-dependent measurements. A comparison of the ARPES data for $h\nu =$~21~eV obtained at $T =$~10~K and 80~K in Figs.~\ref{fig1}(h) does not indicate dramatic changes in band structure. However, an analysis of EDC at the $\bar{\Gamma}$-point in Fig.~\ref{fig1}(h) reveals a temperature-dependence for the state S1. We find that its linewidth increases from $\sim$40~meV (FWHM) at $T =$~40~K and 80~K to $\sim$60~meV at $T =$~10~K. We attribute the enhanced linewidth below $T_N$ to an exchange splitting of the state S1, which is in agreement with recent reports where this splitting into two subbands could be directly resolved \cite{Chen:19,Swatek:19}. From the change in linewidth we estimate an exchange splitting of $\sim$25~meV at $\bar{\Gamma}$, indicating a substantial influence of the Mn magnetic moments on the $p$-derived electronic states near the bottom of the conduction band.

\begin{figure}
\includegraphics[width=\linewidth]{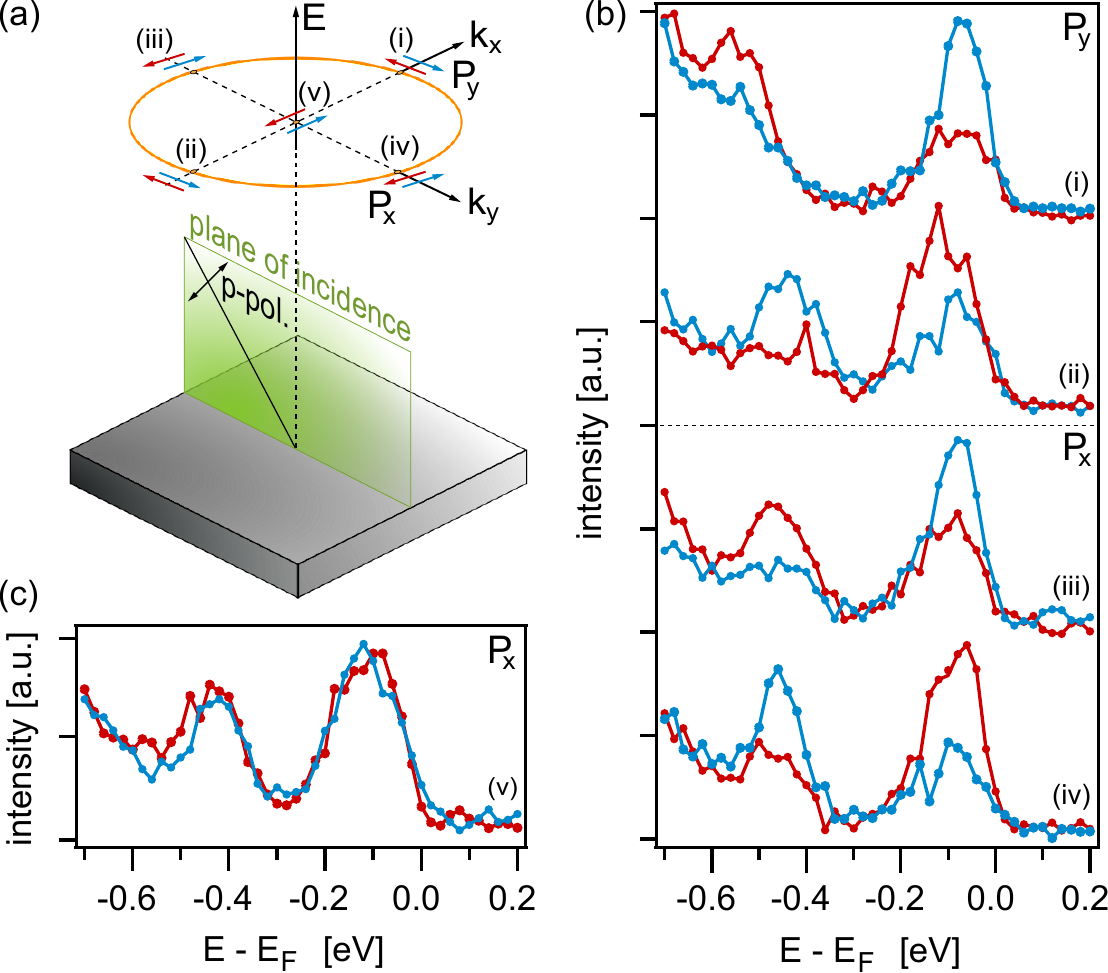}
\caption{(color online) (a) Experimental geometry of the spin-resolved ARPES experiments for MnBi$_2$Te$_4$(0001). Spin-resolved EDC were measured at five different locations in $k_{||}$ space, schematically indicated as (i)-(v) in (a). (b) Spin-resolved EDC obtained at positive and negative wave vectors along $k_x$ and $k_y$. (c)  Spin-resolved EDC at $k_{||}=$~0, showing the absence of spin polarization at $\bar{\Gamma}$. All data sets were acquired at $hv = 20\,$eV and at $T =$~50~K, i.e. in the paramagnetic regime.}
\label{fig3}
\end{figure}

To directly probe the location of the Mn 3$d$ states in the valence band we carried out resonant photoemission experiments at the $L_{2,3}$ absorption edge [Fig.~\ref{fig2}(a)-(b)]. The measured Mn density of states (DOS) shows a main component at ca. 3.8~eV and additional peak at ca. 0.8~eV which is attributed to hybridization between Mn 3$d$ and Te 5$p$ orbitals. The Mn DOS is, thus, very similar to the one of dilute Mn impurities in Sb$_2$Te$_3$ \cite{Islam:18} and GaAs \cite{marco:13}, which suggests a rather local nature of the Mn states in MnBi$_2$Te$_4$. This is in line with the ARPES data set at $hv = 79\,$eV in Fig.~\ref{fig2}(c), where the main Mn compononent in the valence band forms a weakly dispersive feature near 3.8~eV. Based on these data, an accurate theoretical treatment of the localized Mn 3$d$ states and their hybridization with the $p$ bands will be crucial for a comprehensive description of the electronic structure of MnBi$_2$Te$_4$ \cite{marco:13}. 
      
The band inversion in MnBi$_2$Te$_4$(0001) is predicted to arise from the strong spin-orbit interaction in the Bi and Te $p$ states from which the bands near the bulk band gap are derived \cite{otrokov:18,zhang:18,li:18}. We have acquired spin-resolved ARPES data for MnBi$_2$Te$_4$(0001) to confirm the effect of spin-orbit interaction on the surface electronic structure [Fig.~\ref{fig3}]. The measurements were performed at $h\nu =$~20~eV and therefore mainly reflect the state S1 at energies near $E_F$ [see also Figs. S4 an S5 in the supplemental material]. As depicted in Fig.~\ref{fig3}(a) spin-polarized data was obtained in two experimental geometries, namely parallel and perpendicular to the plane of light incidence. In both cases the in-plane spin component perpendicular to the wave vector $k_{||}$ was probed. For both geometries the spin-resolved EDC exhibit a considerable spin-polarization. Moreover, the spin polarization reverses upon changing the sign of $k_{||}$ and it vanishes at $k_{||}= 0$. This observed Rashba-type spin polarization remains robust also for different photon energies [Fig. S4]. 

As the spin-resolved measurements were performed in the paramagnetic regime, the state S1 is expected to be spin-degenerate. Nevertheless, also in the case of spin-degenerate bulk states and surface resonances, strong spin-orbit interaction in combination with the potential step at the surface can induce a local spin density at the surface with a Rashba-type spin polarization in momentum space \cite{Kimura:10,Krasovskii:11}. Due to the surface sensitivity of the photoemission process this local spin density can be reflected in the photoelectron spin polarization. The topological surface state S2 is expected to feature an intrinsic Rashba-type spin polarization \cite{otrokov:18}, which may also contribute to the observed photoelectron spin polarization. Yet, the weak spectral weight of S2 is insufficient to explain the large observed spin polarization alone [Fig. S5]. We therefore attribute the observed Rashba-type spin polarization mainly to strong spin-orbit interaction in the wave function of the spin-degenerate surface resonance S1, in combination with scattering at the surface potential barrier \cite{Kimura:10,Krasovskii:11}. To verify the spin polarization of S2 it will be desirable to perform spin-resolved measurements over a broader photon-energy range.

In summary, our ARPES experiments reveal a complex surface electronic structure of the van der Waals antiferromagnet MnBi$_2$Te$_4$(0001). The results confirm the presence of a surface state in the bulk energy gap. The measurements also provide evidence for the influence of exchange interaction and spin-orbit coupling on the surface electronic structure. In this regard, our findings support the predicted topologically non-trivial nature of MnBi$_2$Te$_4$, in agreement with recent transport studies \cite{Deng:19,Liu:19}. Yet, further work will be required to reveal how the antiferromagnetic order affects the topology of the surface electronic structure and to understand the complex photoemission characteristics of the surface states in dependence of excitation energy. Our results could provide pathways to exploit the interplay of antiferromagnetism and topology in the emerging material class of van der Waals magnets \cite{gong:19}.

\section{Acknowledgments}
We acknowledge financial support from the DFG through SFB1170 'Tocotronics', SFB1143 'Correlated Magnetism', SPP 1666 'Topological insulators', ERA-Chemistry Programm (RU-776/15-1), and the W\"urzburg-Dresden Cluster of Excellence on Complexity and Topology in Quantum Matter -- \textit{ct.qmat} (EXC 2147, project-id 39085490). We also acknowledge the support by Spanish Ministerio de Economia y Competitividad (MINECO Grant No. FIS2016-75862-P), Academic D.I. Mendeleev Fund Program of Tomsk State University (Project No. 8.1.01.2018), the Saint Petersburg State University grant for scientic investigations (Grant No. 15.61.202.2015), and Russian Foundation for Basic Research (Grant No. 18-52-06009). S.M. acknowledges support by the Swiss National Science Foundation (Grant No. P300P2-171221). This research used resources of the Advanced Light Source, which is a DOE Office of Science User Facility under contract no. DE-AC02-05CH11231. We acknowledge Diamond Light Source for access to beamline I05 (proposals No. SI19278 and No. SI22468) that contributed to the results presented here.  Parts of this research were carried out at PETRA III (DESY, Hamburg, Germany) under Proposal No. I-20180510.  This work has been partly performed in the framework of the Nanoscience Foundry and Fine Analysis (NFFA-MIUR, Italy) facility. M.M.O. acknowledges support by the Diputaci\'on Foral de Gipuzkoa ((SAREA
2018 - RED 2018, project no. 2018-CIEN-000025-01).

\bibliographystyle{apsrev}

\begin{thebibliography}{42}
\expandafter\ifx\csname natexlab\endcsname\relax\def\natexlab#1{#1}\fi
\expandafter\ifx\csname bibnamefont\endcsname\relax
  \def\bibnamefont#1{#1}\fi
\expandafter\ifx\csname bibfnamefont\endcsname\relax
  \def\bibfnamefont#1{#1}\fi
\expandafter\ifx\csname citenamefont\endcsname\relax
  \def\citenamefont#1{#1}\fi
\expandafter\ifx\csname url\endcsname\relax
  \def\url#1{\texttt{#1}}\fi
\expandafter\ifx\csname urlprefix\endcsname\relax\def\urlprefix{URL }\fi
\providecommand{\bibinfo}[2]{#2}
\providecommand{\eprint}[2][]{\url{#2}}

\bibitem[{\citenamefont{K\"onig et~al.}(2007)\citenamefont{K\"onig, Wiedmann,
  Br\"une, Roth, Buhmann, Molenkamp, Qi, and Zhang}}]{konig:07.2}
\bibinfo{author}{\bibfnamefont{M.}~\bibnamefont{K\"onig}},
  \bibinfo{author}{\bibfnamefont{S.}~\bibnamefont{Wiedmann}},
  \bibinfo{author}{\bibfnamefont{C.}~\bibnamefont{Br\"une}},
  \bibinfo{author}{\bibfnamefont{A.}~\bibnamefont{Roth}},
  \bibinfo{author}{\bibfnamefont{H.}~\bibnamefont{Buhmann}},
  \bibinfo{author}{\bibfnamefont{L.~W.} \bibnamefont{Molenkamp}},
  \bibinfo{author}{\bibfnamefont{X.}~\bibnamefont{Qi}}, \bibnamefont{and}
  \bibinfo{author}{\bibfnamefont{S.}~\bibnamefont{Zhang}},
  \bibinfo{journal}{Science} \textbf{\bibinfo{volume}{318}},
  \bibinfo{pages}{766} (\bibinfo{year}{2007}).

\bibitem[{\citenamefont{Hasan and Kane}(2010)}]{Hasan:10.11}
\bibinfo{author}{\bibfnamefont{M.~Z.} \bibnamefont{Hasan}} \bibnamefont{and}
  \bibinfo{author}{\bibfnamefont{C.~L.} \bibnamefont{Kane}},
  \bibinfo{journal}{Rev. Mod. Phys.} \textbf{\bibinfo{volume}{82}},
  \bibinfo{pages}{3045} (\bibinfo{year}{2010}).

\bibitem[{\citenamefont{Chang et~al.}(2013)\citenamefont{Chang, Zhang, Feng,
  Shen, Zhang, Guo, Li, Ou, Wei, Wang et~al.}}]{chang:13}
\bibinfo{author}{\bibfnamefont{C.-Z.} \bibnamefont{Chang}},
  \bibinfo{author}{\bibfnamefont{J.}~\bibnamefont{Zhang}},
  \bibinfo{author}{\bibfnamefont{X.}~\bibnamefont{Feng}},
  \bibinfo{author}{\bibfnamefont{J.}~\bibnamefont{Shen}},
  \bibinfo{author}{\bibfnamefont{Z.}~\bibnamefont{Zhang}},
  \bibinfo{author}{\bibfnamefont{M.}~\bibnamefont{Guo}},
  \bibinfo{author}{\bibfnamefont{K.}~\bibnamefont{Li}},
  \bibinfo{author}{\bibfnamefont{Y.}~\bibnamefont{Ou}},
  \bibinfo{author}{\bibfnamefont{P.}~\bibnamefont{Wei}},
  \bibinfo{author}{\bibfnamefont{L.-L.} \bibnamefont{Wang}},
  \bibnamefont{et~al.}, \bibinfo{journal}{Science}
  \textbf{\bibinfo{volume}{340}}, \bibinfo{pages}{167} (\bibinfo{year}{2013}).

\bibitem[{\citenamefont{Mogi et~al.}(2017)\citenamefont{Mogi, Kawamura,
  Yoshimi, Tsukazaki, Kozuka, Shirakawa, Takahashi, Kawasaki, and
  Tokura}}]{mogi:17}
\bibinfo{author}{\bibfnamefont{M.}~\bibnamefont{Mogi}},
  \bibinfo{author}{\bibfnamefont{M.}~\bibnamefont{Kawamura}},
  \bibinfo{author}{\bibfnamefont{R.}~\bibnamefont{Yoshimi}},
  \bibinfo{author}{\bibfnamefont{A.}~\bibnamefont{Tsukazaki}},
  \bibinfo{author}{\bibfnamefont{Y.}~\bibnamefont{Kozuka}},
  \bibinfo{author}{\bibfnamefont{N.}~\bibnamefont{Shirakawa}},
  \bibinfo{author}{\bibfnamefont{K.~S.} \bibnamefont{Takahashi}},
  \bibinfo{author}{\bibfnamefont{M.}~\bibnamefont{Kawasaki}}, \bibnamefont{and}
  \bibinfo{author}{\bibfnamefont{Y.}~\bibnamefont{Tokura}},
  \bibinfo{journal}{Nature Materials} \textbf{\bibinfo{volume}{16}},
  \bibinfo{pages}{516} (\bibinfo{year}{2017}).

\bibitem[{\citenamefont{Xiao et~al.}(2018)\citenamefont{Xiao, Jiang, Shin,
  Wang, Wang, Zhao, Liu, Wu, Chan, Samarth et~al.}}]{Xiao:18}
\bibinfo{author}{\bibfnamefont{D.}~\bibnamefont{Xiao}},
  \bibinfo{author}{\bibfnamefont{J.}~\bibnamefont{Jiang}},
  \bibinfo{author}{\bibfnamefont{J.-H.} \bibnamefont{Shin}},
  \bibinfo{author}{\bibfnamefont{W.}~\bibnamefont{Wang}},
  \bibinfo{author}{\bibfnamefont{F.}~\bibnamefont{Wang}},
  \bibinfo{author}{\bibfnamefont{Y.-F.} \bibnamefont{Zhao}},
  \bibinfo{author}{\bibfnamefont{C.}~\bibnamefont{Liu}},
  \bibinfo{author}{\bibfnamefont{W.}~\bibnamefont{Wu}},
  \bibinfo{author}{\bibfnamefont{M.~H.~W.} \bibnamefont{Chan}},
  \bibinfo{author}{\bibfnamefont{N.}~\bibnamefont{Samarth}},
  \bibnamefont{et~al.}, \bibinfo{journal}{Phys. Rev. Lett.}
  \textbf{\bibinfo{volume}{120}}, \bibinfo{pages}{056801}
  (\bibinfo{year}{2018}).

\bibitem[{\citenamefont{He et~al.}(2017)\citenamefont{He, Pan, Stern, Burks,
  Che, Yin, Wang, Lian, Zhou, Choi et~al.}}]{He:17}
\bibinfo{author}{\bibfnamefont{Q.~L.} \bibnamefont{He}},
  \bibinfo{author}{\bibfnamefont{L.}~\bibnamefont{Pan}},
  \bibinfo{author}{\bibfnamefont{A.~L.} \bibnamefont{Stern}},
  \bibinfo{author}{\bibfnamefont{E.~C.} \bibnamefont{Burks}},
  \bibinfo{author}{\bibfnamefont{X.}~\bibnamefont{Che}},
  \bibinfo{author}{\bibfnamefont{G.}~\bibnamefont{Yin}},
  \bibinfo{author}{\bibfnamefont{J.}~\bibnamefont{Wang}},
  \bibinfo{author}{\bibfnamefont{B.}~\bibnamefont{Lian}},
  \bibinfo{author}{\bibfnamefont{Q.}~\bibnamefont{Zhou}},
  \bibinfo{author}{\bibfnamefont{E.~S.} \bibnamefont{Choi}},
  \bibnamefont{et~al.}, \bibinfo{journal}{Science}
  \textbf{\bibinfo{volume}{357}}, \bibinfo{pages}{294} (\bibinfo{year}{2017}).

\bibitem[{\citenamefont{Mong et~al.}(2010)\citenamefont{Mong, Essin, and
  Moore}}]{Moore:10}
\bibinfo{author}{\bibfnamefont{R.~S.~K.} \bibnamefont{Mong}},
  \bibinfo{author}{\bibfnamefont{A.~M.} \bibnamefont{Essin}}, \bibnamefont{and}
  \bibinfo{author}{\bibfnamefont{J.~E.} \bibnamefont{Moore}},
  \bibinfo{journal}{Phys. Rev. B} \textbf{\bibinfo{volume}{81}},
  \bibinfo{pages}{245209} (\bibinfo{year}{2010}).

\bibitem[{\citenamefont{Chen et~al.}(2010)\citenamefont{Chen, Chu, Analytis,
  Liu, Igarashi, Kuo, Qi, Mo, Moore, Lu et~al.}}]{chen:10}
\bibinfo{author}{\bibfnamefont{Y.~L.} \bibnamefont{Chen}},
  \bibinfo{author}{\bibfnamefont{J.-H.} \bibnamefont{Chu}},
  \bibinfo{author}{\bibfnamefont{J.~G.} \bibnamefont{Analytis}},
  \bibinfo{author}{\bibfnamefont{Z.~K.} \bibnamefont{Liu}},
  \bibinfo{author}{\bibfnamefont{K.}~\bibnamefont{Igarashi}},
  \bibinfo{author}{\bibfnamefont{H.-H.} \bibnamefont{Kuo}},
  \bibinfo{author}{\bibfnamefont{X.~L.} \bibnamefont{Qi}},
  \bibinfo{author}{\bibfnamefont{S.~K.} \bibnamefont{Mo}},
  \bibinfo{author}{\bibfnamefont{R.~G.} \bibnamefont{Moore}},
  \bibinfo{author}{\bibfnamefont{D.~H.} \bibnamefont{Lu}},
  \bibnamefont{et~al.}, \bibinfo{journal}{Science}
  \textbf{\bibinfo{volume}{329}}, \bibinfo{pages}{659} (\bibinfo{year}{2010}).


\bibitem[{\citenamefont{Lee et~al.}(2013)\citenamefont{Lee, Kim, Park, Chung,
  Lim, Seo, and Park}}]{Lee:13}
\bibinfo{author}{\bibfnamefont{D.~S.} \bibnamefont{Lee}},
  \bibinfo{author}{\bibfnamefont{T.-H.} \bibnamefont{Kim}},
  \bibinfo{author}{\bibfnamefont{C.-H.} \bibnamefont{Park}},
  \bibinfo{author}{\bibfnamefont{C.-Y.} \bibnamefont{Chung}},
  \bibinfo{author}{\bibfnamefont{Y.~S.} \bibnamefont{Lim}},
  \bibinfo{author}{\bibfnamefont{W.-S.} \bibnamefont{Seo}}, \bibnamefont{and}
  \bibinfo{author}{\bibfnamefont{H.-H.} \bibnamefont{Park}},
  \bibinfo{journal}{CrystEngComm} \textbf{\bibinfo{volume}{15}},
  \bibinfo{pages}{5532} (\bibinfo{year}{2013}).


\bibitem[{\citenamefont{Eremeev et~al.}(2017)\citenamefont{Eremeev, Otrokov,
  and Chulkov}}]{eremeev:17}
\bibinfo{author}{\bibfnamefont{S.~V.} \bibnamefont{Eremeev}},
  \bibinfo{author}{\bibfnamefont{M.~M.} \bibnamefont{Otrokov}},
  \bibnamefont{and} \bibinfo{author}{\bibfnamefont{E.~V.}
  \bibnamefont{Chulkov}}, \bibinfo{journal}{J. Alloys Compd.}
  \textbf{\bibinfo{volume}{709}}, \bibinfo{pages}{172} (\bibinfo{year}{2017}).

\bibitem[{\citenamefont{Otrokov et~al.}(2018)\citenamefont{Otrokov,
  Klimovskikh, Bentmann, Zeugner, Aliev, Ga{\ss}, Wolter, Koroleva, Estyunin,
  Shikin et~al.}}]{otrokov:18}
\bibinfo{author}{\bibfnamefont{M.~M.} \bibnamefont{Otrokov}},
  \bibinfo{author}{\bibfnamefont{I.~I.} \bibnamefont{Klimovskikh}},
  \bibinfo{author}{\bibfnamefont{H.}~\bibnamefont{Bentmann}},
  \bibinfo{author}{\bibfnamefont{A.}~\bibnamefont{Zeugner}},
  \bibinfo{author}{\bibfnamefont{Z.~S.} \bibnamefont{Aliev}},
  \bibinfo{author}{\bibfnamefont{S.}~\bibnamefont{Ga{\ss}}},
  \bibinfo{author}{\bibfnamefont{A.~U.~B.} \bibnamefont{Wolter}},
  \bibinfo{author}{\bibfnamefont{A.~V.} \bibnamefont{Koroleva}},
  \bibinfo{author}{\bibfnamefont{D.}~\bibnamefont{Estyunin}},
  \bibinfo{author}{\bibfnamefont{A.~M.} \bibnamefont{Shikin}},
  \bibnamefont{et~al.}, \bibinfo{journal}{arXiv:1809.07389 [cond-mat]}
  (\bibinfo{year}{2018}).

\bibitem[{\citenamefont{Zhang et~al.}(2018)\citenamefont{Zhang, Shi, He, Xing,
  Zhang, and Wang}}]{zhang:18}
\bibinfo{author}{\bibfnamefont{D.}~\bibnamefont{Zhang}},
  \bibinfo{author}{\bibfnamefont{M.}~\bibnamefont{Shi}},
  \bibinfo{author}{\bibfnamefont{T.}~\bibnamefont{Zhu}},
  \bibinfo{author}{\bibfnamefont{D.}~\bibnamefont{Xing}},
  \bibinfo{author}{\bibfnamefont{H.}~\bibnamefont{Zhang}}, \bibnamefont{and}
  \bibinfo{author}{\bibfnamefont{J.}~\bibnamefont{Wang}},
  \bibinfo{journal}{Phys. Rev. Lett.}\textbf{\bibinfo{volume}{122}}, \bibinfo{pages}{206401} (\bibinfo{year}{2019}).

\bibitem[{\citenamefont{Li et~al.}(2018)\citenamefont{Li, Li, Du, Wang, Gu,
  Zhang, He, Duan, and Xu}}]{li:18}
\bibinfo{author}{\bibfnamefont{J.}~\bibnamefont{Li}},
  \bibinfo{author}{\bibfnamefont{Y.}~\bibnamefont{Li}},
  \bibinfo{author}{\bibfnamefont{S.}~\bibnamefont{Du}},
  \bibinfo{author}{\bibfnamefont{Z.}~\bibnamefont{Wang}},
  \bibinfo{author}{\bibfnamefont{B.-L.} \bibnamefont{Gu}},
  \bibinfo{author}{\bibfnamefont{S.-C.} \bibnamefont{Zhang}},
  \bibinfo{author}{\bibfnamefont{K.}~\bibnamefont{He}},
  \bibinfo{author}{\bibfnamefont{W.}~\bibnamefont{Duan}}, \bibnamefont{and}
  \bibinfo{author}{\bibfnamefont{Y.}~\bibnamefont{Xu}},
  \bibinfo{journal}{Science Advances} \textbf{\bibinfo{volume}{5}}, \bibinfo{pages}{eaaw5685} (\bibinfo{year}{2019}).

\bibitem[{\citenamefont{Gong et~al.}(2018)\citenamefont{Gong, Guo, Li, Zhu,
  Liao, Liu, Zhang, Gu, Tang, Feng et~al.}}]{gong:18}
\bibinfo{author}{\bibfnamefont{Y.}~\bibnamefont{Gong}},
  \bibinfo{author}{\bibfnamefont{J.}~\bibnamefont{Guo}},
  \bibinfo{author}{\bibfnamefont{J.}~\bibnamefont{Li}},
  \bibinfo{author}{\bibfnamefont{K.}~\bibnamefont{Zhu}},
  \bibinfo{author}{\bibfnamefont{M.}~\bibnamefont{Liao}},
  \bibinfo{author}{\bibfnamefont{X.}~\bibnamefont{Liu}},
  \bibinfo{author}{\bibfnamefont{Q.}~\bibnamefont{Zhang}},
  \bibinfo{author}{\bibfnamefont{L.}~\bibnamefont{Gu}},
  \bibinfo{author}{\bibfnamefont{L.}~\bibnamefont{Tang}},
  \bibinfo{author}{\bibfnamefont{X.}~\bibnamefont{Feng}}, \bibnamefont{et~al.}
  \bibinfo{journal}{Chin. Phys. Lett.} \textbf{\bibinfo{volume}{36}}, \bibinfo{pages}{076801} (\bibinfo{year}{2019}) 

\bibitem[{\citenamefont{Neupane et~al.}(2012)\citenamefont{Neupane, Xu, Wray,
  Petersen, Shankar, Alidoust, Liu, Fedorov, Ji, Allred et~al.}}]{Neupane:12}
\bibinfo{author}{\bibfnamefont{M.}~\bibnamefont{Neupane}},
  \bibinfo{author}{\bibfnamefont{S.-Y.} \bibnamefont{Xu}},
  \bibinfo{author}{\bibfnamefont{L.~A.} \bibnamefont{Wray}},
  \bibinfo{author}{\bibfnamefont{A.}~\bibnamefont{Petersen}},
  \bibinfo{author}{\bibfnamefont{R.}~\bibnamefont{Shankar}},
  \bibinfo{author}{\bibfnamefont{N.}~\bibnamefont{Alidoust}},
  \bibinfo{author}{\bibfnamefont{C.}~\bibnamefont{Liu}},
  \bibinfo{author}{\bibfnamefont{A.}~\bibnamefont{Fedorov}},
  \bibinfo{author}{\bibfnamefont{H.}~\bibnamefont{Ji}},
  \bibinfo{author}{\bibfnamefont{J.~M.} \bibnamefont{Allred}},
  \bibnamefont{et~al.}, \bibinfo{journal}{Phys. Rev. B}
  \textbf{\bibinfo{volume}{85}}, \bibinfo{pages}{235406}
  (\bibinfo{year}{2012}).

\bibitem[{\citenamefont{Okamoto et~al.}(2012)\citenamefont{Okamoto, Kuroda,
  Miyahara, Miyamoto, Okuda, Aliev, Babanly, Amiraslanov, Shimada, Namatame
  et~al.}}]{Okamoto:12}
\bibinfo{author}{\bibfnamefont{K.}~\bibnamefont{Okamoto}},
  \bibinfo{author}{\bibfnamefont{K.}~\bibnamefont{Kuroda}},
  \bibinfo{author}{\bibfnamefont{H.}~\bibnamefont{Miyahara}},
  \bibinfo{author}{\bibfnamefont{K.}~\bibnamefont{Miyamoto}},
  \bibinfo{author}{\bibfnamefont{T.}~\bibnamefont{Okuda}},
  \bibinfo{author}{\bibfnamefont{Z.~S.} \bibnamefont{Aliev}},
  \bibinfo{author}{\bibfnamefont{M.~B.} \bibnamefont{Babanly}},
  \bibinfo{author}{\bibfnamefont{I.~R.} \bibnamefont{Amiraslanov}},
  \bibinfo{author}{\bibfnamefont{K.}~\bibnamefont{Shimada}},
  \bibinfo{author}{\bibfnamefont{H.}~\bibnamefont{Namatame}},
  \bibnamefont{et~al.}, \bibinfo{journal}{Phys. Rev. B}
  \textbf{\bibinfo{volume}{86}}, \bibinfo{pages}{195304}
  (\bibinfo{year}{2012}).

\bibitem[{\citenamefont{Arita et~al.}(2014)\citenamefont{Arita, Sato, Shimada,
  Namatame, Taniguchi, Sasaki, Kitaura, Ohnishi, and Kim}}]{arita:14}
\bibinfo{author}{\bibfnamefont{M.}~\bibnamefont{Arita}},
  \bibinfo{author}{\bibfnamefont{H.}~\bibnamefont{Sato}},
  \bibinfo{author}{\bibfnamefont{K.}~\bibnamefont{Shimada}},
  \bibinfo{author}{\bibfnamefont{H.}~\bibnamefont{Namatame}},
  \bibinfo{author}{\bibfnamefont{M.}~\bibnamefont{Taniguchi}},
  \bibinfo{author}{\bibfnamefont{M.}~\bibnamefont{Sasaki}},
  \bibinfo{author}{\bibfnamefont{M.}~\bibnamefont{Kitaura}},
  \bibinfo{author}{\bibfnamefont{A.}~\bibnamefont{Ohnishi}}, \bibnamefont{and}
  \bibinfo{author}{\bibfnamefont{H.-J.} \bibnamefont{Kim}}, in
  \emph{\bibinfo{booktitle}{Proceedings of the 12th {Asia} {Pacific} {Physics}
  {Conference} ({APPC}12)}} (\bibinfo{publisher}{Journal of the Physical
  Society of Japan}, \bibinfo{year}{2014}), vol.~\bibinfo{volume}{1} of
  \emph{\bibinfo{series}{{JPS} {Conference} {Proceedings}}}.
	
	\bibitem[{\citenamefont{Chen et~al.}(2009)\citenamefont{Chen, Analytis, Chu,
  Liu, Mo, Qi, Zhang, Lu, Dai, Fang et~al.}}]{chen:09}
\bibinfo{author}{\bibfnamefont{Y.~L.} \bibnamefont{Chen}},
  \bibinfo{author}{\bibfnamefont{J.~G.} \bibnamefont{Analytis}},
  \bibinfo{author}{\bibfnamefont{J.-H.} \bibnamefont{Chu}},
  \bibinfo{author}{\bibfnamefont{Z.~K.} \bibnamefont{Liu}},
  \bibinfo{author}{\bibfnamefont{S.-K.} \bibnamefont{Mo}},
  \bibinfo{author}{\bibfnamefont{X.~L.} \bibnamefont{Qi}},
  \bibinfo{author}{\bibfnamefont{H.~J.} \bibnamefont{Zhang}},
  \bibinfo{author}{\bibfnamefont{D.~H.} \bibnamefont{Lu}},
  \bibinfo{author}{\bibfnamefont{X.}~\bibnamefont{Dai}},
  \bibinfo{author}{\bibfnamefont{Z.}~\bibnamefont{Fang}}, \bibnamefont{et~al.},
  \bibinfo{journal}{Science} \textbf{\bibinfo{volume}{325}},
  \bibinfo{pages}{178} (\bibinfo{year}{2009}).

\bibitem[{\citenamefont{Souma et~al.}(2012)\citenamefont{Souma, Eto, Nomura,
  Nakayama, Sato, Takahashi, Segawa, and Ando}}]{Souma:12}
\bibinfo{author}{\bibfnamefont{S.}~\bibnamefont{Souma}},
  \bibinfo{author}{\bibfnamefont{K.}~\bibnamefont{Eto}},
  \bibinfo{author}{\bibfnamefont{M.}~\bibnamefont{Nomura}},
  \bibinfo{author}{\bibfnamefont{K.}~\bibnamefont{Nakayama}},
  \bibinfo{author}{\bibfnamefont{T.}~\bibnamefont{Sato}},
  \bibinfo{author}{\bibfnamefont{T.}~\bibnamefont{Takahashi}},
  \bibinfo{author}{\bibfnamefont{K.}~\bibnamefont{Segawa}}, \bibnamefont{and}
  \bibinfo{author}{\bibfnamefont{Y.}~\bibnamefont{Ando}},
  \bibinfo{journal}{Phys. Rev. Lett.} \textbf{\bibinfo{volume}{108}},
  \bibinfo{pages}{116801} (\bibinfo{year}{2012}).

\bibitem[{\citenamefont{Zeugner et~al.}(2018)\citenamefont{Zeugner, Nietschke,
  Wolter, Ga{\ss}, Vidal, Peixoto, Pohl, Damm, Lubk, Hentrich
  et~al.}}]{Isaeva:18}
\bibinfo{author}{\bibfnamefont{A.}~\bibnamefont{Zeugner}},
  \bibinfo{author}{\bibfnamefont{F.}~\bibnamefont{Nietschke}},
  \bibinfo{author}{\bibfnamefont{A.~U.~B.} \bibnamefont{Wolter}},
  \bibinfo{author}{\bibfnamefont{S.}~\bibnamefont{Ga{\ss}}},
  \bibinfo{author}{\bibfnamefont{R.~C.} \bibnamefont{Vidal}},
  \bibinfo{author}{\bibfnamefont{T.~R.~F.} \bibnamefont{Peixoto}},
  \bibinfo{author}{\bibfnamefont{D.}~\bibnamefont{Pohl}},
  \bibinfo{author}{\bibfnamefont{C.}~\bibnamefont{Damm}},
  \bibinfo{author}{\bibfnamefont{A.}~\bibnamefont{Lubk}},
  \bibinfo{author}{\bibfnamefont{R.}~\bibnamefont{Hentrich}},
  \bibnamefont{et~al.}, \bibinfo{journal}{Chem. Mater.} \textbf{\bibinfo{volume}{31}},
  \bibinfo{pages}{2795} (\bibinfo{year}{2019}).

\bibitem[{\citenamefont{Lee et~al.}(2018)\citenamefont{Lee, Zhu, Wang, Miao,
  Pillsbury, Kempinger, Graf, Alem, Chang, Samarth et~al.}}]{lee:18}
\bibinfo{author}{\bibfnamefont{S.~H.} \bibnamefont{Lee}},
  \bibinfo{author}{\bibfnamefont{Y.}~\bibnamefont{Zhu}},
  \bibinfo{author}{\bibfnamefont{Y.}~\bibnamefont{Wang}},
  \bibinfo{author}{\bibfnamefont{L.}~\bibnamefont{Miao}},
  \bibinfo{author}{\bibfnamefont{T.}~\bibnamefont{Pillsbury}},
  \bibinfo{author}{\bibfnamefont{S.}~\bibnamefont{Kempinger}},
  \bibinfo{author}{\bibfnamefont{D.}~\bibnamefont{Graf}},
  \bibinfo{author}{\bibfnamefont{N.}~\bibnamefont{Alem}},
  \bibinfo{author}{\bibfnamefont{C.-Z.} \bibnamefont{Chang}},
  \bibinfo{author}{\bibfnamefont{N.}~\bibnamefont{Samarth}},
  \bibnamefont{et~al.}, \bibinfo{journal}{arXiv:1812.00339 [cond-mat]}
  (\bibinfo{year}{2018}).
	
		\bibitem{Yan:19} J.-Q. Yan, Q. Zhang, T. Heitmann, Z. Huang, W. D. Wu, D. Vaknin, B. C. Sales, R. J. McQueeney, Phys. Rev. Materials \textbf{3}, 064202 (2019).







\bibitem[{\citenamefont{Rienks et~al.}(2018)\citenamefont{Rienks, Wimmer,
  Mandal, Caha, Růžička, Ney, Steiner, Volobuev, Groiss, Albu
  et~al.}}]{rienks:18}
\bibinfo{author}{\bibfnamefont{E.~D.~L.} \bibnamefont{Rienks}},
  \bibinfo{author}{\bibfnamefont{S.}~\bibnamefont{Wimmer}},
  \bibinfo{author}{\bibfnamefont{P.~S.} \bibnamefont{Mandal}},
  \bibinfo{author}{\bibfnamefont{O.}~\bibnamefont{Caha}},
  \bibinfo{author}{\bibfnamefont{J.}~\bibnamefont{Růžička}},
  \bibinfo{author}{\bibfnamefont{A.}~\bibnamefont{Ney}},
  \bibinfo{author}{\bibfnamefont{H.}~\bibnamefont{Steiner}},
  \bibinfo{author}{\bibfnamefont{V.~V.} \bibnamefont{Volobuev}},
  \bibinfo{author}{\bibfnamefont{H.}~\bibnamefont{Groiss}},
  \bibinfo{author}{\bibfnamefont{M.}~\bibnamefont{Albu}}, \bibnamefont{et~al.}
  (\bibinfo{year}{2018}), \bibinfo{note}{arXiv:1810.06238}.
	
\bibitem[{\citenamefont{Otrokov
  et~al.}(2017{\natexlab{a}})\citenamefont{Otrokov, Menshchikova, Vergniory,
  Rusinov, Vyazovskaya, Koroteev, Bihlmayer, {A. Ernst}, Echenique, Arnau
  et~al.}}]{otrokov:17}
\bibinfo{author}{\bibfnamefont{M.~M.} \bibnamefont{Otrokov}},
  \bibinfo{author}{\bibfnamefont{T.~V.} \bibnamefont{Menshchikova}},
  \bibinfo{author}{\bibfnamefont{M.~G.} \bibnamefont{Vergniory}},
  \bibinfo{author}{\bibfnamefont{I.~P.} \bibnamefont{Rusinov}},
  \bibinfo{author}{\bibfnamefont{A.~Y.} \bibnamefont{Vyazovskaya}},
  \bibinfo{author}{\bibfnamefont{Y.~M.} \bibnamefont{Koroteev}},
  \bibinfo{author}{\bibfnamefont{G.}~\bibnamefont{Bihlmayer}},
  \bibinfo{author}{\bibnamefont{{A. Ernst}}},
  \bibinfo{author}{\bibfnamefont{P.~M.} \bibnamefont{Echenique}},
  \bibinfo{author}{\bibfnamefont{A.}~\bibnamefont{Arnau}},
  \bibnamefont{et~al.}, \bibinfo{journal}{2D Materials}
  \textbf{\bibinfo{volume}{4}}, \bibinfo{pages}{025082}
  (\bibinfo{year}{2017}{\natexlab{a}}).
	
	
\bibitem[{\citenamefont{Otrokov
  et~al.}(2017{\natexlab{b}})\citenamefont{Otrokov, Menshchikova, Rusinov,
  Vergniory, Kuznetsov, and Chulkov}}]{otrokov:17_2}
\bibinfo{author}{\bibfnamefont{M.~M.} \bibnamefont{Otrokov}},
  \bibinfo{author}{\bibfnamefont{T.~V.} \bibnamefont{Menshchikova}},
  \bibinfo{author}{\bibfnamefont{I.~P.} \bibnamefont{Rusinov}},
  \bibinfo{author}{\bibfnamefont{M.~G.} \bibnamefont{Vergniory}},
  \bibinfo{author}{\bibfnamefont{V.~M.} \bibnamefont{Kuznetsov}},
  \bibnamefont{and} \bibinfo{author}{\bibfnamefont{E.~V.}
  \bibnamefont{Chulkov}}, \bibinfo{journal}{JETP Letters}
  \textbf{\bibinfo{volume}{105}}, \bibinfo{pages}{297}
  (\bibinfo{year}{2017}{\natexlab{b}}).
	
	
	

\bibitem[{\citenamefont{Hirahara et~al.}(2017)\citenamefont{Hirahara, Eremeev,
  Shirasawa, Okuyama, Kubo, Nakanishi, Akiyama, Takayama, Hajiri, Ideta
  et~al.}}]{hirahara:17}
\bibinfo{author}{\bibfnamefont{T.}~\bibnamefont{Hirahara}},
  \bibinfo{author}{\bibfnamefont{S.~V.} \bibnamefont{Eremeev}},
  \bibinfo{author}{\bibfnamefont{T.}~\bibnamefont{Shirasawa}},
  \bibinfo{author}{\bibfnamefont{Y.}~\bibnamefont{Okuyama}},
  \bibinfo{author}{\bibfnamefont{T.}~\bibnamefont{Kubo}},
  \bibinfo{author}{\bibfnamefont{R.}~\bibnamefont{Nakanishi}},
  \bibinfo{author}{\bibfnamefont{R.}~\bibnamefont{Akiyama}},
  \bibinfo{author}{\bibfnamefont{A.}~\bibnamefont{Takayama}},
  \bibinfo{author}{\bibfnamefont{T.}~\bibnamefont{Hajiri}},
  \bibinfo{author}{\bibfnamefont{S.-i.} \bibnamefont{Ideta}},
  \bibnamefont{et~al.}, \bibinfo{journal}{Nano Lett.}
  \textbf{\bibinfo{volume}{17}}, \bibinfo{pages}{3493} (\bibinfo{year}{2017}).
	
	
	
	
	
	
	
	
	
	
	
	
	
	
	
	
	






	\bibitem{Otrokov:19} M. M. Otrokov, I. P. Rusinov, M. Blanco-Rey, M. Hoffmann, A. Yu. Vyazovskaya, S. V. Eremeev, A. Ernst, P. M. Echenique, A. Arnau, and E. V. Chulkov. Phys. Rev. Lett. \textbf{122}, 107202 (2019).




	\bibitem{Wang:19} Bo Chen, Fucong Fei, Dongqin Zhang, Bo Zhang, Wanling Liu, Shuai Zhang, Pengdong Wang, Boyuan Wei, Yong Zhang, Zewen Zuo, Jingwen Guo, Qianqian Liu, Zilu Wang, Xuchuan Wu, Junyu Zong, Xuedong Xie, Wang Chen, Zhe Sun, Dawei Shen, Shancai Wang, Yi Zhang, Minhao Zhang, Xuefeng Wang, Fengqi Song, Haijun Zhang, Baigeng Wang, arXiv:1903.09934v1 (2019)



\bibitem{Supp} See Supplemental Material at [URL] for a summary of additional X-ray spectroscopy and ARPES data.

\bibitem[{\citenamefont{Kuiper et~al.}(1993)\citenamefont{Kuiper, Searle,
  Rudolf, Tjeng, and Chen}}]{Kuiper:93}
\bibinfo{author}{\bibfnamefont{P.}~\bibnamefont{Kuiper}},
  \bibinfo{author}{\bibfnamefont{B.~G.} \bibnamefont{Searle}},
  \bibinfo{author}{\bibfnamefont{P.}~\bibnamefont{Rudolf}},
  \bibinfo{author}{\bibfnamefont{L.~H.} \bibnamefont{Tjeng}}, \bibnamefont{and}
  \bibinfo{author}{\bibfnamefont{C.~T.} \bibnamefont{Chen}},
  \bibinfo{journal}{Phys. Rev. Lett.} \textbf{\bibinfo{volume}{70}},
  \bibinfo{pages}{1549} (\bibinfo{year}{1993}).

\bibitem[{\citenamefont{Alders et~al.}(1998)\citenamefont{Alders, Tjeng, Voogt,
  Hibma, Sawatzky, Chen, Vogel, Sacchi, and Iacobucci}}]{Alders:98}
\bibinfo{author}{\bibfnamefont{D.}~\bibnamefont{Alders}},
  \bibinfo{author}{\bibfnamefont{L.~H.} \bibnamefont{Tjeng}},
  \bibinfo{author}{\bibfnamefont{F.~C.} \bibnamefont{Voogt}},
  \bibinfo{author}{\bibfnamefont{T.}~\bibnamefont{Hibma}},
  \bibinfo{author}{\bibfnamefont{G.~A.} \bibnamefont{Sawatzky}},
  \bibinfo{author}{\bibfnamefont{C.~T.} \bibnamefont{Chen}},
  \bibinfo{author}{\bibfnamefont{J.}~\bibnamefont{Vogel}},
  \bibinfo{author}{\bibfnamefont{M.}~\bibnamefont{Sacchi}}, \bibnamefont{and}
  \bibinfo{author}{\bibfnamefont{S.}~\bibnamefont{Iacobucci}},
  \bibinfo{journal}{Phys. Rev. B} \textbf{\bibinfo{volume}{57}},
  \bibinfo{pages}{11623} (\bibinfo{year}{1998}).

\bibitem[{\citenamefont{Abbate et~al.}(1992)\citenamefont{Abbate, Goedkoop,
  Groot, Grioni, Fuggle, Hofmann, Petersen, and Sacchi}}]{abbate:92}
\bibinfo{author}{\bibfnamefont{M.}~\bibnamefont{Abbate}},
  \bibinfo{author}{\bibfnamefont{J.~B.} \bibnamefont{Goedkoop}},
  \bibinfo{author}{\bibfnamefont{F.~M. F.~d.} \bibnamefont{Groot}},
  \bibinfo{author}{\bibfnamefont{M.}~\bibnamefont{Grioni}},
  \bibinfo{author}{\bibfnamefont{J.~C.} \bibnamefont{Fuggle}},
  \bibinfo{author}{\bibfnamefont{S.}~\bibnamefont{Hofmann}},
  \bibinfo{author}{\bibfnamefont{H.}~\bibnamefont{Petersen}}, \bibnamefont{and}
  \bibinfo{author}{\bibfnamefont{M.}~\bibnamefont{Sacchi}},
  \bibinfo{journal}{Surface and Interface Analysis}
  \textbf{\bibinfo{volume}{18}}, \bibinfo{pages}{65} (\bibinfo{year}{1992}).



\bibitem[{\citenamefont{Kurata and Colliex}(1993)}]{Kurata:93}
\bibinfo{author}{\bibfnamefont{H.}~\bibnamefont{Kurata}} \bibnamefont{and}
  \bibinfo{author}{\bibfnamefont{C.}~\bibnamefont{Colliex}},
  \bibinfo{journal}{Phys. Rev. B} \textbf{\bibinfo{volume}{48}},
  \bibinfo{pages}{2102} (\bibinfo{year}{1993}).

\bibitem[{\citenamefont{Qiao et~al.}(2013)\citenamefont{Qiao, Chin, Harris,
  Yan, and Yang}}]{qiao:13}
\bibinfo{author}{\bibfnamefont{R.}~\bibnamefont{Qiao}},
  \bibinfo{author}{\bibfnamefont{T.}~\bibnamefont{Chin}},
  \bibinfo{author}{\bibfnamefont{S.~J.} \bibnamefont{Harris}},
  \bibinfo{author}{\bibfnamefont{S.}~\bibnamefont{Yan}}, \bibnamefont{and}
  \bibinfo{author}{\bibfnamefont{W.}~\bibnamefont{Yang}},
  \bibinfo{journal}{Current Applied Physics} \textbf{\bibinfo{volume}{13}},
  \bibinfo{pages}{544} (\bibinfo{year}{2013}).
	
	
	
	
	
	
	
\bibitem[{\citenamefont{Sanchez-Barriga
  et~al.}(2016)\citenamefont{Sanchez-Barriga, Varykhalov, Springholz, Steiner,
  Kirchschlager, Bauer, Caha, Schierle, Weschke, \"Unal et~al.}}]{sanchez:16}
\bibinfo{author}{\bibfnamefont{J.}~\bibnamefont{Sanchez-Barriga}},
  \bibinfo{author}{\bibfnamefont{A.}~\bibnamefont{Varykhalov}},
  \bibinfo{author}{\bibfnamefont{G.}~\bibnamefont{Springholz}},
  \bibinfo{author}{\bibfnamefont{H.}~\bibnamefont{Steiner}},
  \bibinfo{author}{\bibfnamefont{R.}~\bibnamefont{Kirchschlager}},
  \bibinfo{author}{\bibfnamefont{G.}~\bibnamefont{Bauer}},
  \bibinfo{author}{\bibfnamefont{O.}~\bibnamefont{Caha}},
  \bibinfo{author}{\bibfnamefont{E.}~\bibnamefont{Schierle}},
  \bibinfo{author}{\bibfnamefont{E.}~\bibnamefont{Weschke}},
  \bibinfo{author}{\bibfnamefont{A.~A.} \bibnamefont{\"Unal}},
  \bibnamefont{et~al.}, \bibinfo{journal}{Nature Communications}
  \textbf{\bibinfo{volume}{7}}, \bibinfo{pages}{10559} (\bibinfo{year}{2016}).



\bibitem[{\citenamefont{Xia et~al.}(2009)\citenamefont{Xia, Qian, Hsieh, Wray,
  Pal, Lin, Bansil, Grauer, Hor, Cava et~al.}}]{xia:09}
\bibinfo{author}{\bibfnamefont{Y.}~\bibnamefont{Xia}},
  \bibinfo{author}{\bibfnamefont{D.}~\bibnamefont{Qian}},
  \bibinfo{author}{\bibfnamefont{D.}~\bibnamefont{Hsieh}},
  \bibinfo{author}{\bibfnamefont{L.}~\bibnamefont{Wray}},
  \bibinfo{author}{\bibfnamefont{A.}~\bibnamefont{Pal}},
  \bibinfo{author}{\bibfnamefont{H.}~\bibnamefont{Lin}},
  \bibinfo{author}{\bibfnamefont{A.}~\bibnamefont{Bansil}},
  \bibinfo{author}{\bibfnamefont{D.}~\bibnamefont{Grauer}},
  \bibinfo{author}{\bibfnamefont{Y.~S.} \bibnamefont{Hor}},
  \bibinfo{author}{\bibfnamefont{R.~J.} \bibnamefont{Cava}},
  \bibnamefont{et~al.}, \bibinfo{journal}{Nature Physics}
  \textbf{\bibinfo{volume}{5}}, \bibinfo{pages}{398} (\bibinfo{year}{2009}).
	
	
\bibitem{Hao:19}	Yu-Jie Hao, Pengfei Liu, Yue Feng, Xiao-Ming Ma, Eike F. Schwier, Masashi Arita, Shiv Kumar, Chaowei Hu, Rui'e Lu, Meng Zeng, Yuan Wang, Zhanyang Hao, Hongyi Sun, Ke Zhang, Jiawei Mei, Ni Ni, Liusuo Wu, Kenya Shimada, Chaoyu Chen, Qihang Liu, Chang Liu, arXiv:1907.03722 (2019).

\bibitem{Chen:19} Y. J. Chen, L. X. Xu, J. H. Li, Y. W. Li, C. F. Zhang, H. Li, Y. Wu, A. J. Liang, C. Chen, S. W. Jung, C. Cacho, H. Y. Wang, Y. H. Mao, S. Liu, M. X. Wang, Y. F. Guo, Y. Xu, Z. K. Liu, L. X. Yang, Y. L. Chen, 	arXiv:1907.05119 (2019).

\bibitem{Swatek:19} Przemyslaw Swatek, Yun Wu, Lin-Lin Wang, Kyungchan Lee, Benjamin Schrunk, Jiaqiang Yan, Adam Kaminski, arXiv:1907.09596 (2019).

\bibitem{Vidal:19} Raphael C. Vidal, Alexander Zeugner, Jorge I. Facio, Rajyavardhan Ray, M. Hossein Haghighi, Anja U. B. Wolter, Laura T. Corredor Bohorquez, Federico Caglieris, Simon Moser, Tim Figgemeier, Thiago R. F. Peixoto, Hari Babu Vasili, Manuel Valvidares, Sungwon Jung, Cephise Cacho, Alexey Alfonsov, Kavita Mehlawat, Vladislav Kataev, Christian Hess, Manuel Richter, Bernd B\"uchner, Jeroen van den Brink, Michael Ruck, Friedrich Reinert, Hendrik Bentmann, Anna Isaeva, arXiv:1906.08394 (2019).
	










\bibitem[{\citenamefont{Islam et~al.}(2018)\citenamefont{Islam, Canali,
  Pertsova, Balatsky, Mahatha, Carbone, Barla, Kokh, Tereshchenko, Jim\'enez
  et~al.}}]{Islam:18}
\bibinfo{author}{\bibfnamefont{M.~F.} \bibnamefont{Islam}},
  \bibinfo{author}{\bibfnamefont{C.~M.} \bibnamefont{Canali}},
  \bibinfo{author}{\bibfnamefont{A.}~\bibnamefont{Pertsova}},
  \bibinfo{author}{\bibfnamefont{A.}~\bibnamefont{Balatsky}},
  \bibinfo{author}{\bibfnamefont{S.~K.} \bibnamefont{Mahatha}},
  \bibinfo{author}{\bibfnamefont{C.}~\bibnamefont{Carbone}},
  \bibinfo{author}{\bibfnamefont{A.}~\bibnamefont{Barla}},
  \bibinfo{author}{\bibfnamefont{K.~A.} \bibnamefont{Kokh}},
  \bibinfo{author}{\bibfnamefont{O.~E.} \bibnamefont{Tereshchenko}},
  \bibinfo{author}{\bibfnamefont{E.}~\bibnamefont{Jim\'enez}},
  \bibnamefont{et~al.}, \bibinfo{journal}{Phys. Rev. B}
  \textbf{\bibinfo{volume}{97}}, \bibinfo{pages}{155429}
  (\bibinfo{year}{2018}).
	
	
\bibitem[{\citenamefont{Marco et~al.}(2013)\citenamefont{Marco, Thunström,
  Katsnelson, Sadowski, Karlsson, Lebègue, Kanski, and Eriksson}}]{marco:13}
\bibinfo{author}{\bibfnamefont{I.~D.} \bibnamefont{Marco}},
  \bibinfo{author}{\bibfnamefont{P.}~\bibnamefont{Thunström}},
  \bibinfo{author}{\bibfnamefont{M.~I.} \bibnamefont{Katsnelson}},
  \bibinfo{author}{\bibfnamefont{J.}~\bibnamefont{Sadowski}},
  \bibinfo{author}{\bibfnamefont{K.}~\bibnamefont{Karlsson}},
  \bibinfo{author}{\bibfnamefont{S.}~\bibnamefont{Lebègue}},
  \bibinfo{author}{\bibfnamefont{J.}~\bibnamefont{Kanski}}, \bibnamefont{and}
  \bibinfo{author}{\bibfnamefont{O.}~\bibnamefont{Eriksson}},
  \bibinfo{journal}{Nature Communications} \textbf{\bibinfo{volume}{4}},
  \bibinfo{pages}{2645} (\bibinfo{year}{2013}).




		\bibitem{Kimura:10} A. Kimura, E. E. Krasovskii, R. Nishimura, K. Miyamoto, T. Kadono, K. Kanomaru, E. V. Chulkov, G. Bihlmayer, K. Shimada, H. Namatame, and M. Taniguchi, Phys. Rev. Lett. \textbf{105}, 076804 (2010).
	
	\bibitem{Krasovskii:11} E. E. Krasovskii and E. V. Chulkov, Phys. Rev. B \textbf{83}, 155401 (2011).
	
	\bibitem{Deng:19} Yujun Deng, Yijun Yu, Meng Zhu Shi, Jing Wang, Xian Hui Chen, Yuanbo Zhang, arXiv:1904.11468 (2019).
	
	\bibitem{Liu:19} Chang Liu, Yongchao Wang, Hao Li, Yang Wu, Yaoxin Li, Jiaheng Li, Ke He, Yong Xu, Jinsong Zhang, Yayu Wang, arXiv:1905.00715 (2019).

\bibitem[{\citenamefont{Gong and Zhang}(2019)}]{gong:19}
\bibinfo{author}{\bibfnamefont{C.}~\bibnamefont{Gong}} \bibnamefont{and}
  \bibinfo{author}{\bibfnamefont{X.}~\bibnamefont{Zhang}},
  \bibinfo{journal}{Science} \textbf{\bibinfo{volume}{363}},
  \bibinfo{pages}{eaav4450} (\bibinfo{year}{2019}).

\end{thebibliography}

\end{document}